\documentclass[11pt]{article}
\usepackage[dvips]{graphicx}
\begin{document}
\title{Shock structures in time averaged patterns for the Kuramoto-Sivashinsky equation}
\author{Hidetsugu Sakaguchi\\
Department of Applied Science for Electronics and Materials,\\ Interdisciplinary Graduate School of Engineering Sciences,\\
 Kyushu University, Kasuga, Fukuoka 816-8580, Japan}
\maketitle
\begin{abstract}
The Kuramoto-Sivashinsky equation with  
fixed boundary conditions is numerically studied. 
Shocklike structures appear in the time-averaged patterns 
for some parameter range of the boundary values.  Effective diffusion constant is 
estimated from the relation of the width and the height of the
shock structures.  
 \\
PACS numbers:05.45.-a, 47.54.+r
\end{abstract}
\newpage
An ordered structure can be obtained by time averages of disordered spatiotemporal patterns. Such ordered structures are experimentally found in Faraday waves, rotating thermal convection and electroconvection [1-3]. 
These ordered structures seem to come from the boundary effect.
The Kuramoto-Sivashinsky equation is one of the simplest partial differential equations exhibiting spatiotemporal chaos. Average patterns of spatiotemporal chaos in the Kuramoto-Sivashinsky equation were studied in [4]. 
We study average patterns of the Kuramoto-Sivashinsky equation with 
different boundary conditions.
The equation in one dimension has the form
\begin{equation}
h_t=-h_{xx}-h_{xxxx}+\frac{1}{2}(h_x)^2,
\end{equation}
where $h=h(x,t)$ is a real function, $x\in [0,L]$, and the subscripts stand for derivatives.  An equivalent equation is obtained for $u=h_x$ as 
\begin{equation}
u_t=-u_{xx}-u_{xxxx}+uu_{x}.
\end{equation}
Equation (2) possesses the Galilean symmetry$(x\rightarrow x+vt, u\rightarrow u+v)$.  The Kuramoto-Sivashinsky equation exhibits spatiotemporal chaos and it is conjectured that the 
large-scale properties of the equation are described by the stochastic Burgers equation [5-8].  
\begin{equation}
u_t=D u_{xx}+\lambda uu_x+\xi,
\end{equation}
where $D>0$, $\xi(x,t)$ denotes the noise term, and the coefficient $\lambda$ of the nonlinear term is expected to be 1 owing to the same Galilean symmetry as the Kuramoto-Sivashinsky equation. 
The Fourier transform $\xi_k(t)$ of $\xi(x,t)$ satisfies 
\[<\xi_k(t)\xi_{k'}(t')>=Tk^2\delta_{k,k'}\delta(t-t').\]
Characteristic solutions to the Burgers equation without the last noise term 
are the shock solutions:
\[u(x)=A\tanh(\kappa x),\]
where $\kappa=\lambda A/(2D)$.  
We have performed numerical simulations to 
seek for some shocklike structures in large-scale averaged patterns for 
the Kuramoto-Sivashinsky equation. 
We have used a simple real-space scheme obtained by replacing spatial 
and temporal derivatives by finite differences. The spatial discretization step is $\delta x=0.25$ and the temporal discretization step $\delta t=2.5\times 10^{-4}$, and the system size $L=400$.
We have considered fixed boundary conditions, that is,
\[u(0,t)=-U,\;u(L,t)=U,\;u_x(0,t)=u_x(L,t)=0,\]
where $\pm U$ are the fixed values at the boundaries and $U$ is changed as a 
parameter. The initial condition is 
\begin{eqnarray}
u(x,0)&=&-U+Ux/(2l),\;\;\;\;\;\;\;\;{\rm for}\;0<x<l\nonumber\\
&=&U(x-L/2)/(L-2l),\;{\rm for}\;l<x<L-l\nonumber\\
&=&U+U(x-L)/(2l),\;{\rm for}\;L-l<x<L\nonumber\\
\end{eqnarray}
where $l=4$ is assumed. The simulation is performed until $t=t_2=1500$ and the 
temporal average is calculated as 
\[\bar{u}(x)=(1/N)\sum_{n=1}^{N}u(x,t_1+n\Delta t)\]
where $t_1=500$, $\Delta t=50\delta t$ and $N=(t_2-t_1)/\Delta t$.

Figure 1 displays snapshot patterns at $t=1500$ and the averaged 
pattern. In the central region of the averaged pattern, $\bar{u}$ takes nearly 0 and  
the mean slope of $\bar u$ is very small for $U<1.3$. 
This type of averaged patterns is obtained for $U=0$ in [4,6].

A stationary shocklike pattern appears at $U=1.65$.
Since the solution is stationary in time, the averaged pattern and the snapshot pattern overlap in Fig. 1(c).
This stationary solution can be understood from the 
absolute stability of a constant solution as follows.
A constant solution $u(x,t)=U$ is convectively unstable, but 
it becomes absolutely stable for $U>1.622$.
The stability of the constant solution $u(x,t)=U$ is determined from the 
linear dispersion relation for the perturbation $\delta u(x,t)\sim u_ke^{ikx-i\omega t}$ around $u(x,t)=U$ [9]. The linear dispersion relation is written as 
\begin{equation}
\omega(k)=-Uk+i(k^2-k^4).
\end{equation}  
The solution is absolutely stable under the condition that  $d\omega/dk=-U+2i(k-k^3)=0$ and ${\rm Im} \omega(k)=-U{\rm Im} k+{\rm Re} (k^2-k^4)<0$ 
where $k$ is generally a complex number. 
The transition occurs at $U=U_c$ where $U_c$ satisfies  
\[
27U_c^4-68U_c^2-8=0,\]
that is, $U_c=\{(34+\sqrt{34^2+8\cdot 27})/27\}^{1/2}\sim 1.622$.  
The convectively unstable but absolutely stable solution is observed in our system, since the  
fixed boundary conditions are assumed and no disturbance is created at the boundaries.  A solution which connects the two 
absolutely stable solutions $u(x,t)=U$ and $u(x,t)=-U$ appears 
as a shocklike structure with oscillating tails in Fig. 1(c). 

Another type of average patterns appears for $1.3<U<1.6$. 
Two nearly flat regions are observed and a shocklike structure connecting the two 
flat regions appears at $U=1.45$ as shown in Fig.1(b).
The averaged value of $u(x,t)$ in the flat region is not $\pm U$ and $u(x,t)$ exhibits 
spatiotemporal chaos around the averaged value.  
We have assumed the form of the averaged pattern as $A\tanh(\kappa x)$ and  
estimated the value of $A$ and $\kappa$. Figure 2 displays the averaged patterns and the corresponding profile $\bar{u}(x)=A\tanh(\kappa x)$ at $U=1.33$ and $1.45$.
As $U$ is increased, $A$ is increased and $\kappa$ is decreased. 
We have observed this shocklike structure in the time-averaged pattern even if the initial value of $u(x,0)$ is a random number between -0.02 and 0.02. 
We have checked up that the shocklike structure appears 
in the time-averaged pattern for a larger system with $L=3200$. 
It is fairly robust, but we are not sure that the shocklike structure 
is maintained in an infinite size system.

Figure 3 displays the height $A$ of the shocklike structure as a function of 
$U$. The diamonds denote simulation results for $L=400$ and the crosses denote simulation results for $L=1600$.  The shock amplitude $A$ seems to be independent of the system size for sufficiently large $L$. The height $A$ seems to increases from zero continuously, however, 
the transition is not clear, since the spatial fluctuation of the averaged pattern is large near $U\sim 1.3$. Figure 4 displays the effective diffusion constant $D=\lambda A/(2\kappa)$ at $L=400$ (denoted by diamonds) and $L=1600$ (denoted by crosses), where $\lambda=1$ is assumed. 
In our simulations $A,\kappa$ and $D$ depends on $U$. We have plotted 
a relation of $D(U)$ v.s. $\kappa(U)$ in Fig.4.  As $\kappa(U)$ is decreased, $D(U)$ seems to increase. It might be interpreted that the effective 
diffusion constant tends to increase, as the length scale is longer.
Sneppen et al. estimated the effective diffusion constant $D$ and $\lambda$ 
as $D\sim 10.5$ and $\lambda=2.32$ from the analysis of the interface width 
$<(h(x,t)-<h>)^2>$ [8].
Our effective diffusion constant seems to approach about 10 as $\kappa\rightarrow 0$.
This result may be consistent with the result of Sneppen et al., although  they consider a very large system and  
the effective coefficient $\lambda$ of the nonlinear term is not  1 in their analysis.

In summary, we have numerically studied the Kuramoto-Sivashinsky equation 
with fixed boundary conditions.  At both the boundaries, $u(x,t)$ are fixed to $\pm U$.  
If the $U$ is smaller than a critical value, the averaged pattern takes 
nearly zero. If $U$ is larger than another critical value, a stationary 
shock pattern with oscillating tails appears owing to the absolute stability.
We have found a shocklike structure of the spatiotemporal chaos between the two  critical values. We do not understand well the fairly stable structure 
in the spatiotemporal chaos, but it may be partly related to the weakness 
of the absolute instability. We are not sure  whether the shocklike 
structures are due to finite-size effects. 
     
In our finite-size simulations, the time averaged patterns take  
fairly definite nonzero amplitudes $\pm A$. 
The time averaged pattern is approximated with the shock solution 
of the Burgers equation and the effective diffusion constant is estimated. 
The effective diffusion constant  depends on the length scale of the shock width and it seems to approach about 10 for a very long scale.

This work was supported in part by the Grant-in-Aid for Scientific 
Research No.11837014 from the Ministry of Education, Science, Sports and Culture.

\newpage

{\bf Figure captions}\\
Fig. 1. Snapshot patterns (solid lines) $u(x,t)$ at $t=1500$ and the time averaged patterns (dashed lines) $\bar{u}(x)$ at $U=1.25$ (a), 1.45 (b) and 1.65 (c). \\
Fig. 2. Time averaged patterns (solid lines) and their fitting with $\bar{u}=A\tanh(\kappa x)$ (dashed lines) at 
$U=1.33$ (a) and $1.45$  (b).\\
Fig. 3. Relation of the height $A$ v.s. $U$ for the time averaged patterns at $L=400$ (denoted by diamonds) and $L=1600$ (denoted by crosses).\\
Fig. 4. Relation of the effective diffusion constant $D$ v.s. $\kappa$ for 
the time averaged patterns at $L=400$ (denoted by diamonds) and $L=1600$ (denoted by crosses).

\begin{thebibliography}{99}
\bibitem{rf:1} B.~J.~Gluckman, C.~B.~Arnold, and J.~P.~Gollub, Phys. Rev. E {\bf51}, 1128 (1995). 
\bibitem{rf:2} L.~Ning, Y.~Hu, R.~E.~Ecke, and G.~Ahlers, Phys. Rev. Lett. {\bf 71}, 2216 (1996).
\bibitem{rf:3} S.~Rudroff and I.~Rehberg, Phys. Rev. E {\bf 55}, 2742 (1997).
\bibitem{rf:4} V.~M.~Egu\'iluz, P.~Alstr\o m, E.~Hern\'andez-Garc\'ia, and O.~Piro, Phys. Rev. E {\bf 59}, 2822 (1999).
\bibitem{rf:5} V.~Yakhot, Phys. Rev. A {\bf 24}, 642 (1981).
\bibitem{rf:6} S.~Zaleski and L.~Lallemand, J. Phys. Lett. (Paris) {\bf 24}, L793 (1985).
\bibitem{rf:7} S.~Zaleski, Physica D {\bf 34}, 427 (1989).
\bibitem{rf:8} K.~Sneppen, J.~Krug, M.~H.~Jensen, C.~Jayaprakash, and T.~Bohr, Phys. Rev. A {\bf 46}, R7351 (1992).
\bibitem{rf:9} R.~J.~Deissler, Physica D {\bf 25}, 233 (1987). 
\end{thebibliography}
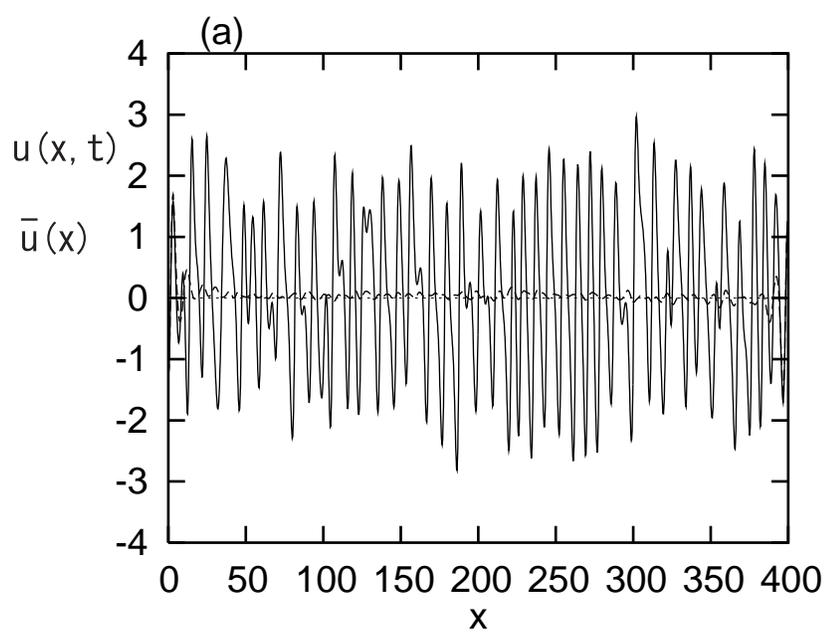
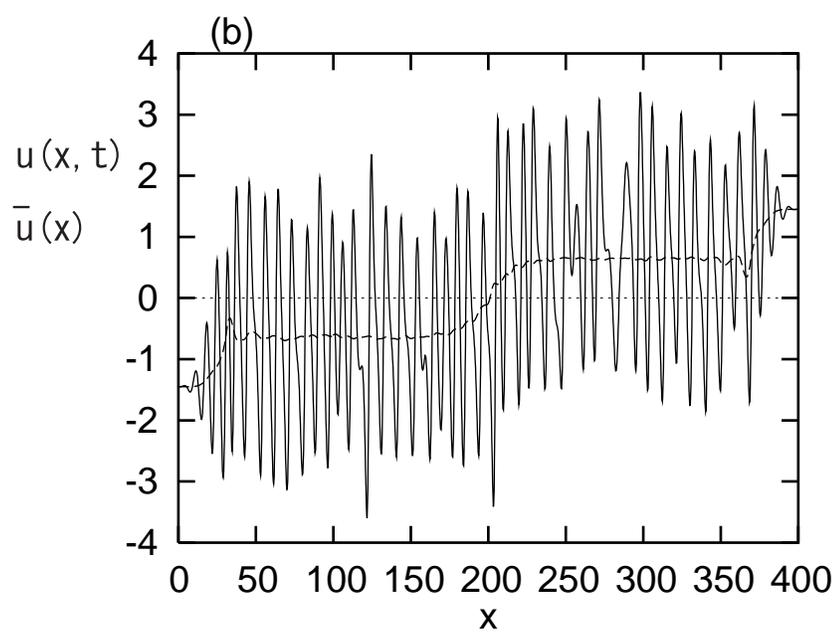
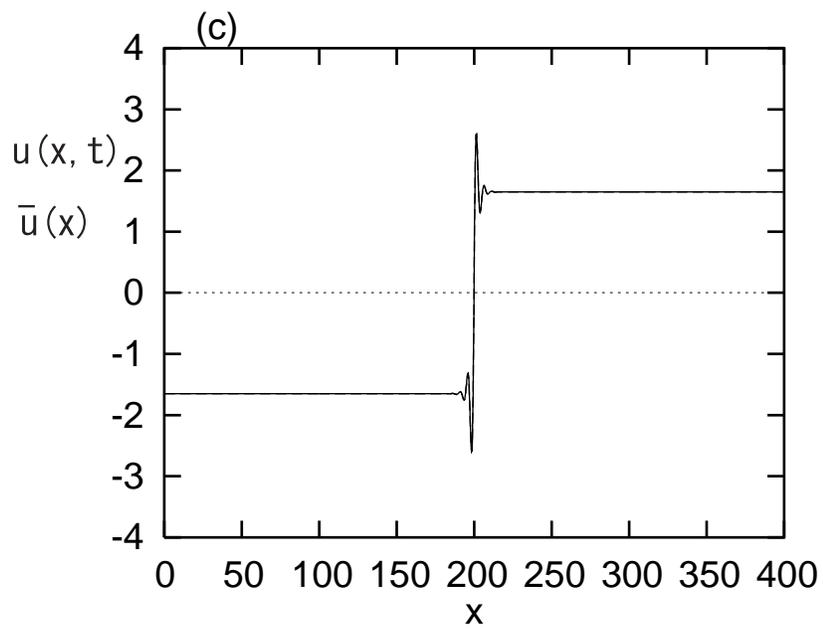
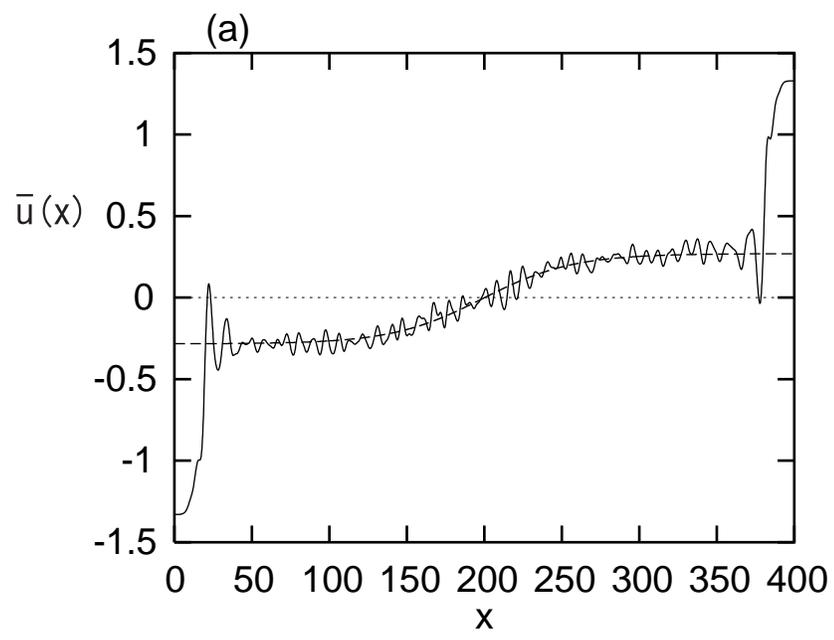
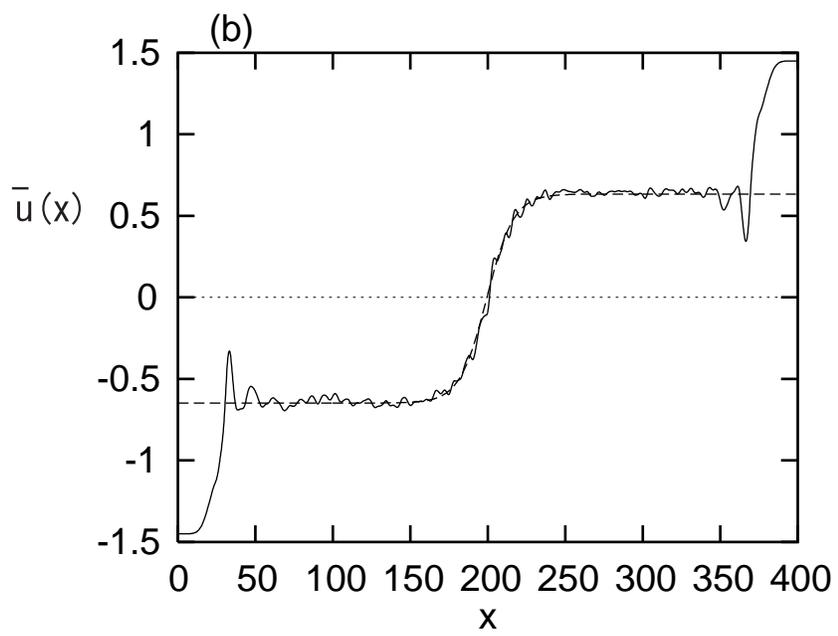
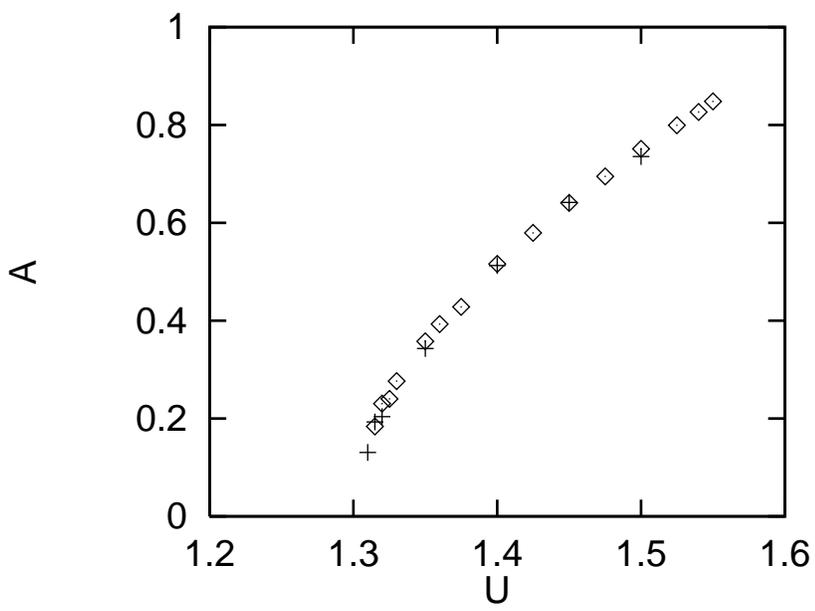
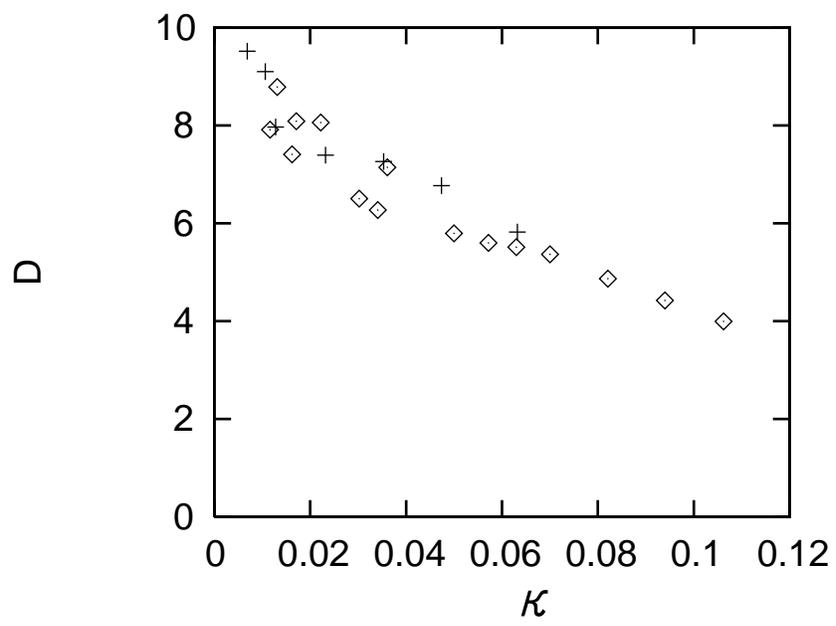
\end{document}